\documentclass[12pt]{article}%,a4]{article}
\usepackage{setspace}
\usepackage{amssymb}
\usepackage{amsmath,amssymb}
\usepackage{amsmath}
\usepackage[T1]{fontenc}
\usepackage[bf,small,tableposition=top]{caption}
\usepackage{subfig}
\usepackage{epsfig}
\usepackage{amsthm}
\usepackage{lineno}
\usepackage[figuresright]{rotating}
\usepackage{enumerate}
\usepackage{authblk}
\usepackage[T1]{fontenc}
\usepackage[utf8]{inputenc}
\usepackage{graphics}
\usepackage{graphicx}
\usepackage{multirow}
\usepackage{lscape}
\usepackage{graphicx}
\usepackage{caption}
\usepackage[dvipsnames]{xcolor}
\usepackage{mathrsfs}
\usepackage{epstopdf}
\usepackage{hyperref}
\usepackage{textcomp}
\usepackage{listings}
\usepackage{amsmath}
\usepackage{amsthm}
\usepackage{float}

\usepackage{tipa}
\usepackage{verbatim}
\usepackage{diagbox}
\usepackage{multirow,bigdelim,dcolumn,booktabs}

\theoremstyle{definition}

\theoremstyle{remark}

\usepackage{multirow}
\begin{document}
	
	\vspace{1cm}
	\begin{center}
		\textbf{ Composite Lognormal - $T$ regression models with varying threshold and its insurance application }
	\end{center}
	
	\begin{center}
		Girish Aradhye$^\dagger$, Deepesh Bhati$^\dagger$\footnote{Corresponding author: deepesh.bhati@curaj.ac.in}, 	George Tzougas$^*$  \\
		$^\dagger$Department of Statistics, Central University of Rajasthan, Ajmer, India.\\
		$^*$Maxwell Institute for Mathematical Sciences and Department of Actuarial Mathematics and Statistics, Heriot-Watt University, Edinburgh, EH14 4AS, United Kingdom. 
	\end{center}

	\begin{abstract}
		Composite probability models have shown very promising results for modeling claim severity data comprised of small, moderate and large losses. In this paper, we introduce three classes of parametric composite regression models with varying threshold. We consider the Lognormal distribution for the head and the Burr, the Stoppa and the generalized log-Moyal (GlogM) distributions for the tail part of the composite family. Further, the Mode-Matching procedure has been utilized for the composition of the two  densities. To capture the heterogeneous behavior of the  policyholder’s characteristics, covariates are introduced into the scale parameter of the tail distribution. Finally, the applicability of the proposed  models has been shown using a real-world insurance data set.  \\ 
		
	\end{abstract}

	\section{Introduction}
One of the most distinguishing characteristics of the claim severity data set is that it contains small and intermediate values with  high frequency, as well as a few significant values with  low frequency. Moreover, these data are uni-modal and heavily right-skewed \cite{2,4}. When the data spans over a wide range of magnitude, choosing a suitable single probability distribution that gives a good fit to both the small and large regions of the empirical distribution of claims becomes difficult. This necessitated novel methods. Some popular methods used/developed in this direction are  (i) transformation of  random variables (r.v.) \cite{29,1,16,12}, mixture of two or more distributions \cite{28,20}, (ii) method of compounding \cite{23} and (iii) method of composition of distributions \cite{10,24,9,15,25,21}. \cite{21} proposed different composite models by considering Burr, Loglogistic, Paralogistic and Generalized Pareto distribution for the tail of the data using  unrestricted mixing weights and truncated densities before and after the threshold. Typically, continuity and differentiability conditions are imposed  at threshold  to ensure the smoothness of the resultant density, which results in the reduction of model parameters \cite{10,24}.      \\
\indent To ease the process of making composite density differentiable as well as the parameter reduction, \cite{8} introduced another composite model by matching the two families of distribution at modal value and proposed Lognormal-Stoppa and Weibull-Stoppa models. This approach replaces the differentiability condition at the threshold point. As indicated in \cite{8}, in the Mode-Matching procedure, reduction in parameters can be easier than  the computation of second-order derivatives in the traditional differentiability condition. Thereafter, \cite{8}and \cite{6} proposed composite models based on the Mode-Matching method, where they  considered different choices for the tail and head distributions. \\
\indent Often the claim severity data is influenced by the individual characteristics of the policyholder. These characteristics are defined in terms of covariates and are available along with claim severity data. \cite{14} pointed out that, for many lines of insurance business, actuarial data often exhibit more extreme tail behavior than suggested by the normal distribution. Modeling such heavy-tailed data sets using available covariates through regression analysis can be a practical tool for claim severity analysis. Modeling the claim severity data using composite models, the covariates which are readily available along with claim amount are recently being  considered. Some recent work on composite regression models includes \cite{14} , \cite{18} , \cite{13}  and \cite{19}. \cite{14} proposed composite regression models and analyzed Singapore car claims data. They advocated the use of two composite regression models comprising three components to capture both fat-tail data behavior and policyholder heterogeneity by introducing the available covariates in the scale parameter of the second and third component of the composite regression models. \cite{18} studied the composite Lognormal-Pareto Type II regression model and analyzed household budget data via the Particle Swarm Optimization algorithm to estimate model parameters. \cite{13} presented a mixture composite regression model to tackle the challenges in claim severity modeling such as multimodality and heavy tail behaviour by utilizing covariates information in the body and tail of the distribution. In the above regression applications to composite models, classical composite models developed by imposing classical continuity and differentiability conditions at threshold are used. In all the above mentioned work, a threshold is a fixed parameter. However, in general, considering the fixed threshold point for a heterogeneous group of individuals is quite impractical \cite{22}. As mentioned in  \cite{22}, "estimating the threshold point together with the other parameters accounts for threshold uncertainty but assuming a unique threshold point applying to all the claims may appear quite unrealistic". In line with this, \cite{19} extended the actuarial literature on composite regression models by studying the GBII composite regression model with varying threshold point where both the head and tail part of the composite regression models contains the GBII distribution. \cite{19} incorporated the covariates information in the location parameter of the tail part of the composite regression model so that the policyholder heterogeneity among the individuals in the tail parts of loss data can be captured by the GBII composite regression model. The parameters involved in the GBII composite regression model are relatively more as compared to the other two-parameter heavy-tailed distributions with no parameter regularization which could lead to an over-fitting problem and make model interpretation difficult \cite{27}. Hence in this work, we propose a composite regression model with varying threshold, which has  a limited number of parameters. This parsimony will reduce the computational burden of the estimation procedure and also give a reasonably good fit to the claim severity data where covariates impact various parts of the claim severity data sets with straightforward model interpretation. \\
\indent In this paper we introduce three new families of composite regression models by using the Mode-Matching technique. The proposed models are suitable for modeling the positive skewness and heavy-tailed response in the presence of individual characteristics of claimants. We name the proposed composite regression model as the "composite Lognormal-$T$ regression model". We consider the  Lognormal (LN) distribution for the head part of the composite regression models and three well known heavy-tailed distributions namely the Burr (B), the Stoppa (S) \cite{26} and generalized log-Moyal (GlogM) \cite{7} distribution for the tail part ($T$) of composite regression models. The analytical expressions for the mode of the above mentioned  heavy-tailed distributions are available in the close form, which enable us to reduce the model parameter. We incorporate the covariates in the scale parameter of the tail distribution of the proposed family of composite regression models. In case of the Mode-Matching composite model, the mode of the tail part of the composite density is the threshold point from where the two non-overlapping densities (head and tail part) separate. The covariate-dependent scale parameter is involved in the analytical expression of the mode of the tail part of the composite model, which results in the varying threshold point concerning individual policyholders’ characteristics. This enables us to study the impact of available covariates on the modal part as well as the heavy-tailed part of the claim severity data set.

The rest of the paper is structured as follows. In section \ref{MMF}, the modeling framework involved in the generation of composite models using  Mode-Matching (MM) technique is presented. Section \ref{RAMM} presents the detailed discussion on the composite Lognormal-$T$  regression model where $T$ stands for the distribution at the tail part of the composite regression models. Computational aspects involved in the fitting of composite Lognormal-$T$  regression model are presented in section \ref{MECA}. Empirical investigation and modeling results using a real life insurance data set are demonstrated in Section \ref{RDA}. Finally, some concluding remarks can be found in section \ref{CONC}.
	
	\section{Modeling framework: Mode-Matching Technique }\label{MMF}
	\cite{8} gave a viewpoint of compositing two densities at the mode of the data. The densities on both sides of the mode are taken into account depending on how quickly the probability declines from the mode value. As a result, the resulting distribution provides more balanced support for the data. The head distribution of the composite model is used up to mode value (which is to be estimated from data) thereafter the truncated family of tail distribution is considered. The probability density function (pdf) of the composite distribution obtained from the Mode-Matching technique is given as,  
	\begin{equation} 
		f(y)=
		\begin{cases} \label{uwm}
			r f^{*}_{H}(y), &  \text{for} \qquad 0< y \le y_{mo}\\
			(1-r)f^{*}_{T}(y), & \text{for} \qquad y_{mo}< y <\infty.
		\end{cases}
	\end{equation}
	where  $r \in [0,1]$ is the unrestricted mixing weight associated with the head part of the composite density,  and  $f^{*}_{H}(y)=\frac{f_{H}(y)}{F_{H}(y_{mo})}$ and $f^{*}_{T}(y)=\frac{f_{T}(y)}{1-F_{T}(y_{mo})}$. Where $f_{H}(y)$ and $f_{T}(y)$ are the pdf of the head and tail component of composite model, $F_{H}(y_{mo})$ and $F_{T}(y_{mo})$ are the cumulative distribution function (cdf) of the head and tail component of composite model evaluated at the mode of the tail distribution $y_{mo}$ which is also a threshold point. We denote the mode of the head and the tail distribution by $y^{H}_{mo}$ and $y^{T}_{mo}$ respectively. The conditions for the  Mode-Matching procedure are
	\begin{align} \label{mmc1}
		y^{H}_{m0}&=y^{T}_{m0},\\ \label{mmc2}
		rf^{*}_{H}(y^{H}_{m0})&=(1-r)f^{*}_{T}(y^{T}_{m0}).
	\end{align} 
	($\ref{mmc2}$) implies the continuity condition is satisfied. Equality in ($\ref{mmc1}$) allows us to drop the $\textit{'H'}$ and $\textit{'T'}$ labels, and gives the weight parameter $r$ as
	\begin{equation}
		r=\frac{f_{T}(y_{mo})F_{H}(y_{mo})}{f_{T}(y_{mo})F_{H}(y_{mo})+f_{H}(y_{mo})(1-F_{T}(y_{mo}))},
	\end{equation}
	condition ($\ref{mmc1}$) surpass the differentiability condition, as  for uni-model distribution, the derivative of density at the mode is zero. In this paper we make use of Mode-Matching technique to generate proposed composite regression models.

	\section{Regression Application for Mode-Matching Composite Lognormal - $T$  Models}\label{RAMM}
	\indent The pdf of composite Lognormal-$T$   model can be written as
	\begin{equation}\label{crmpdf}
		f(y)=
		\begin{cases}
			r_{LN,T}\frac{\exp\{-\frac{1}{2}\left(\frac{\log(y)-\mu}{\sigma}\right)^2\}}{\sqrt{2\pi}\sigma y\Phi\left(\frac{\log(y_{mo})-\mu}{\sigma}\right)}, & \text{for} \quad  0< y \le y_{mo} \\\\
			(1-r_{LN,T})f_{T}^{*}(y;\alpha,\beta), &\text{for} \quad  y_{mo} < y<\infty.
		\end{cases}
	\end{equation} 
	
	The cdf of composite Lognormal-$T$   model may be written as
	\begin{equation}\label{crmcdf}
		F(y)=
		\begin{cases}
			r_{LN,T}\frac{\Phi\left(\frac{\log(y)-\mu}{\sigma}\right)}{\Phi\left(\frac{\log(y_{mo})-\mu}{\sigma}\right)}, &\text{for} \quad 0 < y \le y_{mo}\\
			r_{LN,T}+(1-r_{LN,T}) \frac{F_{T}(y)-F_{T}(y_{mo})}{1-F_{T}(y_{mo})}, & \text{for} \quad  y_{mo} < y < \infty.
		\end{cases}
	\end{equation}
	Where $r_{LN,T} \in [0,1]$, $-\infty < \mu < \infty$, $\sigma > 0$, $\alpha > 0$, threshold point $y_{mo}$ > 0  and scale parameter $\beta > 0$. $r_{LN,T}$ is the mixing weight associated with the right truncated Lognormal density of the composite Lognormal-$T$   model having Lognormal distribution for the head part and $T$ distribution at the tail part. $f_{T}^{*}(y;\alpha,\beta)=\frac{f_{T}(y)}{1-F_{T}(y_{mo})}$ is the adequate left truncated density of the tail part of the composite Lognormal-$T$  model truncated at the threshold point $y_{mo}$. We propose to introduce the available significant covariates in the scale parameter of the composite Lognormal-$T$ regression models which  allow for the estimation of risk behaviour of the individual policyholder.
	
	The regression structure for the scale parameter $\beta$ of the composite Lognormal-$T$  models can be employed as follows
	\begin{align} \label{link}
		\underline{\beta}=\exp(\underline{\gamma}^{\top}\underline{\text{x}} ),
	\end{align}     
	where $\underline{\text{x}}$ is covariates information vector which are assumed to be column vectors having dimension $p \times 1$, with $\underline{\gamma} = (\gamma_{1}, \gamma_{2},\cdots,\gamma_{p})$ are the corresponding regression coefficients. To retrain the non negative support of the scale parameter $\beta$, we use log-link function to link various covariates to $\beta$. The mode of the tail distribution ($T$) of the composite Lognormal-$T$ regression model ($y^{T}_{mo}$) which is also a threshold point contains the covariate dependent scale parameter $\underline{\beta}$. Which intern shows the dependency of threshold point on available covariates. Imposing the continuity condition given in  \ref{mmc2} at the threshold point $y_{mo}$, the expression for the covariate dependent mixing weight i.e. $r_{LN, T}(\bold{x_i};\boldsymbol{\gamma})$ can be written as 
	
	\begin{equation}
		r_{LN, T}(\bold{x_i};\boldsymbol{\gamma})=\frac{f_{T}(y_{mo}(\bold{x_i};\boldsymbol{\gamma}))\Phi\left(\frac{\log(y_{mo}(\bold{x_i};\boldsymbol{\gamma}))-\mu}{\sigma}\right)}{f_{T}(y_{mo}(\bold{x_i};\boldsymbol{\gamma}))\Phi\left(\frac{\log(y_{mo}(\bold{x_i};\boldsymbol{\gamma}))-\mu}{\sigma}\right)+\frac{\exp\{-\frac{1}{2}\left(\frac{\log(y_{mo}(\bold{x_i};\boldsymbol{\gamma}))-\mu}{\sigma}\right)^2\}}{\sqrt{2\pi}\sigma y_{mo}(\bold{x_i};\boldsymbol{\gamma}) } (1-F_{T}(y_{mo}(\bold{x_i};\boldsymbol{\gamma})))}.
	\end{equation}
	
	We use following three distributions for modelling  the tail part of composite Lognormal-$T$  regression models.
	\begin{enumerate}
		\item Burr distribution: \\
		$$f_{B}(y)=\frac{\delta \alpha y^{(\alpha-1)} \beta^{\alpha}}{[(y\beta)^{\alpha}+1]^{(\delta+1)}}  \quad \text{and} \quad  F_{B}(y)=1-\left[\frac{1}{(y \beta )^{\alpha}+1}\right]^{\delta},$$\\
		$\text{for}\quad y > 0, \alpha >0, \beta > 0 , \delta > 0$.
		
		\item GlogM distribution:\\ $$f_{GlogM}(y)=\frac{\beta^\frac{1}{2\alpha}\exp\{-\frac{1}{2}\left(\frac{\beta}{y}\right)^\frac{1}{\alpha}\}}{\sqrt{2\pi}\alpha y^{\frac{1}{2\alpha}+1}} \quad \text{and} \quad F_{GlogM}(y)=1-\left[\text{erf}\left(\frac{1}{\sqrt{2}}\left(\frac{\beta}{y}\right)^{\frac{1}{2\alpha}}\right)\right], $$\\
		\text{for}\quad $y > 0$, $\alpha >0$, $\beta > 0$.

		\item Stoppa distribution: \\ $$f_{S}(y)=\alpha \delta \beta^{\delta}y^{-(\delta+1)} \left[1-\left(\frac{y}{\beta}\right)^{-\delta}\right]^{\alpha-1}  \quad \text{and} \quad F_{S}(y)=\left[1-\left(\frac{y }{\beta}\right)^{-\delta}\right]^{\alpha},$$\\
		\text{for}\quad $y > \beta$, $\alpha >0$, $\beta > 0$ , $\delta > 0$.
		
	\end{enumerate}
	The detailed procedure involved in the generation of composite Lognormal-$T$  regression models for the three parametric distributions for the tail distribution is given in the following section.
	
	\subsection{Mode-Matching Composite Lognormal - Burr  Regression Model}
	
	Let $Y$ be the r.v. follows a composite Lognormal - Burr model with pdf
	\begin{equation}
		f(y)=
		\begin{cases}
			r_{LN, B}\frac{\exp\{-\frac{1}{2}\left(\frac{\log(y)-\mu}{\sigma}\right)^2\}}{\sqrt{2\pi}\sigma y\Phi\left(\frac{\log(y_{mo})-\mu}{\sigma}\right)}, & \text{for} \quad  0< y \le y_{mo} \\\\
			(1-r_{LN, B}) \frac{\delta \alpha y^{(\alpha-1)} \beta^{\alpha} [(y\beta)^{\alpha}+1]^{-(\delta+1)}}{[(y \beta )^{\alpha}+1]^{-\delta}},   &\text{for} \quad  y_{mo} < y<\infty.
		\end{cases}
	\end{equation} 
	with $\mu \in \mathbb{R}$, $\sigma > 0$, the scale parameter $\beta > 0$, $\alpha > 0$, $\delta > 0$, threshold point $y_{mo}$ > 0, $r_{LN, B} \in [0,1]$ be the mixing weight of composite model which constitutes Lognormal head and Burr tail and $\Phi(.)$ denotes the cdf of the standard normal distribution. Here the scale parameter $\beta$ contains the  covariates information  as follows.
	\begin{align} \label{link3}
		\underline{\beta}=\exp(\underline{\gamma}^{\top}\underline{\text{x}} ),
	\end{align}     
	where $\underline{\text{x}}$ is covariates information vector with dimension $p \times 1$, with $\underline{\gamma} = (\gamma_{1}, \gamma_{2},\cdots,\gamma_{p})$ are the corresponding regression coefficients.\\
	After imposing continuity condition  at covariate dependent threshold point $y_{mo}(\bold{x_i};\boldsymbol{\gamma})$ given in  (\ref{mmc2}), we get the 
	
	\begin{equation}
		r_{LN, B}(\bold{x_i};\boldsymbol{\gamma})=\frac{f_{B}(y_{mo}(\bold{x_i};\boldsymbol{\gamma}))\Phi\left(\frac{\log(y_{mo}(\bold{x_i};\boldsymbol{\gamma}))-\mu}{\sigma}\right)}{f_{B}(y_{mo}(\bold{x_i};\boldsymbol{\gamma}))\Phi\left(\frac{\log(y_{mo}(\bold{x_i};\boldsymbol{\gamma}))-\mu}{\sigma}\right)+\frac{\exp\{-\frac{1}{2}\left(\frac{\log(y_{mo}(\bold{x_i};\boldsymbol{\gamma}))-\mu}{\sigma}\right)^2\}}{\sqrt{2\pi}\sigma y_{mo}(\bold{x_i};\boldsymbol{\gamma}) } (1-F_{B}(y_{mo}(\bold{x_i};\boldsymbol{\gamma})))},
	\end{equation}
	where $f_{B}(y_{mo}(\bold{x_i};\boldsymbol{\gamma}))$ and $F_{B}(y_{mo}(\bold{x_i};\boldsymbol{\gamma}))$ are the pdf and cdf of the Burr distribution respectively evaluated at the threshold point $y_{mo}(\bold{x_i};\boldsymbol{\gamma})$. The analytical expressions for the mode of the Lognormal distribution and the Burr distribution is given by \\
	\begin{equation*}
		y^{LN}_{mo}=\text{exp}(\mu-\sigma^{2})\quad \text{and} \quad
		y^{B}_{mo}=\frac{1}{\underline{\beta}} \left(\frac{\alpha-1}{\delta \alpha +1}\right)^{\frac{1}{\alpha}} , \quad  \alpha > 1.
	\end{equation*}
	One can easily see that, mode of the Burr distribution contains the scale parameter $\underline{\beta}$. In case of composite Lognormal-Burr  regression model, the threshold point is the mode of Burr distribution ($y^{B}_{mo}$). The regression setting in (\ref{link3}) permits the scale parameter $\underline{\beta}$ to be linked with available covariates which results in the varying threshold point according to individual peculiarities of the policyholder.  
	Using (\ref{mmc1}), the reduced parameter $\mu$ can be written as 
	\begin{align}\label{muvalue3}
		\text{exp}(\mu-\sigma^{2})= \frac{1}{\underline{\beta}} \left(\frac{\alpha-1}{\delta \alpha +1}\right)^{\frac{1}{\alpha}}  & \quad \implies 	\mu=\sigma^{2}+\log\left(\frac{1}{\underline{\beta}} \left(\frac{\alpha-1}{\delta \alpha +1}\right)^{\frac{1}{\alpha}}\right), \quad \alpha > 1.
	\end{align} 
	Eq. (\ref{muvalue3}) impose additional constraint on $\alpha$ as $\alpha > 1$ to define a positive modal value for the Burr distribution.

	\subsection{Mode-Matching Composite Lognormal - GlogM  Regression Model}
	
	Let $Y$ be the r.v. follows a composite Lognormal - GlogM   model with pdf
	\begin{equation}
		f(y)=
		\begin{cases}
			r_{LN, GlogM}\frac{\exp\{-\frac{1}{2}\left(\frac{\log(y)-\mu}{\sigma}\right)^2\}}{\sqrt{2\pi}\sigma y\Phi\left(\frac{\log(y_{mo})-\mu}{\sigma}\right)}, & \text{for} \quad  0< y \le y_{mo} \\\\
			(1-r_{LN, GlogM})\frac{\beta^\frac{1}{2\alpha}\exp\{-\frac{1}{2}\left(\frac{\beta}{y}\right)^\frac{1}{\alpha}\}}{\sqrt{2\pi}\alpha y^{\frac{1}{2\alpha}+1}\left(\text{erf}\left(\frac{1}{\sqrt{2}}\left(\frac{\beta}{y_{mo}}\right)^{\frac{1}{2\alpha}}\right)\right)}, &\text{for} \quad  y_{mo} < y<\infty.
		\end{cases}
	\end{equation} 
	Where $r_{LN, GlogM} \in [0,1]$ is the mixing weight of the composite model containing Lognormal distribution for head and GlogM distribution for tail, $-\infty < \mu < \infty$, $\sigma > 0$, $\alpha > 0$ and the scale parameter $\beta > 0$, threshold point $y_{mo}$ > 0 and $\Phi(.)$ denotes the cdf of the standard normal distribution. Here the scale parameter $\beta$ can linked to the various covariates as follows.
	\begin{align} \label{link1}
		\underline{\beta}=\exp(\underline{\gamma}^{\top}\underline{\text{x}} ),
	\end{align}     
	where $\underline{\text{x}}$ is covariates information vector which are assumed to be column vectors having dimension $p \times 1$, with $\underline{\gamma} = (\gamma_{1}, \gamma_{2},\cdots,\gamma_{p})$ are the corresponding regression coefficients.\\
	After imposing continuity condition at varying threshold point $y_{mo}(\bold{x_i};\boldsymbol{\gamma})$ given in  (\ref{mmc2}), we get 
	\begin{equation}
		r_{LN, GlogM}(\bold{x_i};\boldsymbol{\gamma})=\frac{f_{GlogM}(y_{mo}(\bold{x_i};\boldsymbol{\gamma}))\Phi\left(\frac{\log(y_{mo}(\bold{x_i};\boldsymbol{\gamma}))-\mu}{\sigma}\right)}{f_{GlogM}(y_{mo}(\bold{x_i};\boldsymbol{\gamma}))\Phi\left(\frac{\log(y_{mo}(\bold{x_i};\boldsymbol{\gamma}))-\mu}{\sigma}\right)+\frac{\exp\{-\frac{1}{2}\left(\frac{\log(y_{mo}(\bold{x_i};\boldsymbol{\gamma}))-\mu}{\sigma}\right)^2\}}{\sqrt{2\pi}\sigma y_{mo}(\bold{x_i};\boldsymbol{\gamma}) } (1-F_{GlogM}(y_{mo}(\bold{x_i};\boldsymbol{\gamma})))},
	\end{equation}
	where $f_{GlogM}(y_{mo}(\bold{x_i};\boldsymbol{\gamma}))$ and $F_{GlogM}(y_{mo}(\bold{x_i};\boldsymbol{\gamma}))$ are the pdf and cdf of the GlogM distribution respectively evaluated at the threshold point $y_{mo}(\bold{x_i};\boldsymbol{\gamma})$. The analytical expressions for the mode of the Lognormal distribution and the GlogM distribution is given by \\
	\begin{equation*}
		y^{LN}_{mo}=\text{exp}(\mu-\sigma^{2})\quad \text{and} \quad
		y^{GlogM}_{mo}=\frac{\underline{\beta}}{(1+2\alpha)^{\alpha}}.
	\end{equation*}
	It is point to be noted that, the covariate dependent scale parameter $\underline{\beta}$ is involved in the expression of mode the GlogM distribution. In case of composite Lognormal-GlogM  regression model, the threshold point is the mode of GlogM distribution ($y^{GlogM}_{mo}$). Which in turn shows the dependency of threshold point on the available covariates.  
	Using (\ref{mmc1}), the reduction in the parameter will took place as
	\begin{align}\label{muvalue}
		\text{exp}(\mu-\sigma^{2})=\frac{\underline{\beta}}{(1+2\alpha)^{\alpha}} & \quad \implies 	\mu=\sigma^{2}+\log\left(\frac{\underline{\beta}}{(1+2\alpha)^{\alpha}}\right).
	\end{align}

	\subsection{Mode-Matching Composite Lognormal - Stoppa  Regression Model}
	
	Let $Y$ be the r.v. follows a composite Lognormal - Stoppa   model with pdf
	\begin{equation}
		f(y)=
		\begin{cases}
			r_{LN, S}\frac{\exp\{-\frac{1}{2}\left(\frac{\log(y)-\mu}{\sigma}\right)^2\}}{\sqrt{2\pi}\sigma y\Phi\left(\frac{\log(y_{mo})-\mu}{\sigma}\right)}, & \text{for} \quad  0< y \le y_{mo} \\\\
			(1-r_{LN, S})\frac{\alpha \delta \beta^{\delta}y^{-(\delta+1)} \left[1-\left(\frac{y}{\beta}\right)^{-\delta}\right]^{\alpha-1} }{1-\left[1-\left(\frac{y_{mo} }{\beta}\right)^{-\delta}\right]^{\alpha}},    &\text{for} \quad  y_{mo} < y<\infty.
		\end{cases}
	\end{equation} 
	with $\mu \in \mathbb{R}$, $\sigma > 0$, $\alpha > 1$, $\delta > 0$, scale parameter $\beta > 0$, threshold point $y_{mo}$ > 0,  $r_{LN, S} \in [0,1]$ be the mixing weight of the composite model having Lognormal head and Stoppa tail and $\Phi(.)$ denotes the cdf of the standard normal distribution. Here the scale parameter $\beta$ contains the  covariates information  as follows.
	\begin{align} \label{link2}
		\underline{\beta}=\exp(\underline{\gamma}^{\top}\underline{\text{x}} ),
	\end{align}     
	where $\underline{\text{x}}$ is covariates information vector with dimension $p \times 1$, with $\underline{\gamma} = (\gamma_{1}, \gamma_{2},\cdots,\gamma_{p})$ are the corresponding regression coefficients.\\
	After imposing continuity condition at covariate dependent threshold point $y_{mo}(\bold{x_i};\boldsymbol{\gamma})$ given in equation (\ref{mmc2}), we get 
	\begin{equation}
		r_{LN, S}(\bold{x_i};\boldsymbol{\gamma})=\frac{f_{S}(y_{mo}(\bold{x_i};\boldsymbol{\gamma}))\Phi\left(\frac{\log(y_{mo}(\bold{x_i};\boldsymbol{\gamma}))-\mu}{\sigma}\right)}{f_{S}(y_{mo}(\bold{x_i};\boldsymbol{\gamma}))\Phi\left(\frac{\log(y_{mo}(\bold{x_i};\boldsymbol{\gamma}))-\mu}{\sigma}\right)+\frac{\exp\{-\frac{1}{2}\left(\frac{\log(y_{mo}(\bold{x_i};\boldsymbol{\gamma}))-\mu}{\sigma}\right)^2\}}{\sqrt{2\pi}\sigma y_{mo}(\bold{x_i};\boldsymbol{\gamma}) } (1-F_{S}(y_{mo}(\bold{x_i};\boldsymbol{\gamma})))},
	\end{equation}
	where $f_{S}(y_{mo}(\bold{x_i};\boldsymbol{\gamma}))$ and $F_{S}(y_{mo}(\bold{x_i};\boldsymbol{\gamma}))$ are the pdf and cdf of the Stoppa distribution respectively evaluated at the threshold point $y_{mo}(\bold{x_i};\boldsymbol{\gamma})$. The expressions for the mode of the Lognormal distribution and the Stoppa distribution is given by \\
	\begin{equation*}
		y^{LN}_{mo}=\text{exp}(\mu-\sigma^{2})\quad \text{and} \quad
		y^{S}_{mo}=\underline{\beta} \left(\frac{1+\alpha \delta}{1+\delta}\right)^{\frac{1}{\delta}}, \quad \alpha > 1.
	\end{equation*}
	The regression setting employed in (\ref{link2}) makes the scale parameter $\underline{\beta}$ covariate dependent. The threshold point $y^{S}_{mo}$ i.e. the mode of Stoppa distribution contains the covariate dependent scale parameter $\underline{\beta}$ which results in the varying threshold across individual policy holders risk behaviour. The parametric relation for the parameter $\mu$ can be written using (\ref{mmc1}) as follows.
	\begin{align}\label{muvalue}
		\text{exp}(\mu-\sigma^{2})= \underline{\beta} \left(\frac{1+\alpha \delta}{1+\delta}\right)^{\frac{1}{\delta}} & \quad \implies 	\mu=\sigma^{2}+\log\left(\underline{\beta} \left(\frac{1+\alpha \delta}{1+\delta}\right)^{\frac{1}{\delta}}\right), \quad \alpha > 1.
	\end{align}
	Note that the additional constraint $\alpha > 1$ is needed for the existence of  mode of the Stoppa distribution.

	\section{Parameter Estimation and Computational Aspects}\label{MECA}
	Let $Y_{1}, Y_{2},\dots, Y_{n}$ be a random sample follows a Mode-Matching composite Lognormal-$T$  model described in the  (\ref{crmpdf}). Here $\boldsymbol{\theta}$ is the parameter vector for the composite Lognormal-$T$ model and $\boldsymbol{\gamma}$ is the vector of regression coefficients associated with the covariate vector $\bold{x_i}$ available along with each $\bold{y_i}$. Let $\boldsymbol{\Theta}=(\boldsymbol{\theta},\boldsymbol{\gamma}$) be the vector of model parameters. We utilize the maximum likelihood (ML) estimation procedure to estimate the model parameters. The goal of maximum likelihood estimation procedure is to find the values for parameters which maximize 
	\begin{equation*}
		l(\boldsymbol{\Theta}|y_i)= \sum_{i=1}^{n}\text{ln}(f(y_{i}|\boldsymbol{\Theta}))
	\end{equation*}
	
	\begin{align*} \nonumber
		& l(\boldsymbol{\Theta}|y_{i})= \sum_{i=1}^{n}\text{ln} \left[	r_{LN,T}(\bold{x_i};\boldsymbol{\gamma}) \frac{\exp\{-\frac{1}{2}\left(\frac{\log(y_{i})-\mu(\bold{x_i};\boldsymbol{\gamma})}{\sigma}\right)^2\}}{\sqrt{2\pi}\sigma y_{i}\Phi\left(\frac{\log(y_{mo}(\bold{x_i};\boldsymbol{\gamma}))-\mu(\bold{x_i};\boldsymbol{\gamma})}{\sigma}\right)} \mathbb{I}[y_{i} < y_{mo}(\bold{x_i};\boldsymbol{\gamma})] \right.\\
		&  \left. \qquad +  	(1-r_{LN,T}(\bold{x_i};\boldsymbol{\gamma}))\frac{f_{T}(y_i)}{1-F_{T}(y_{mo}(\bold{x_i};\boldsymbol{\gamma}))} \mathbb{I}[y_{i} \ge y_{mo}(\bold{x_i};\boldsymbol{\gamma})]\right] 
	\end{align*}

	\begin{align} \nonumber
		l(\boldsymbol{\Theta}|y_{i})=& \sum_{i=1}^{n} \left[	\ln (r_{LN,T}(\bold{x_i};\boldsymbol{\gamma}))\mathbb{I}[y_{i} < y_{mo}(\bold{x_i};\boldsymbol{\gamma})]+\ln \left(\frac{\exp\{-\frac{1}{2}\left(\frac{\log(y_{i})-\mu(\bold{x_i};\boldsymbol{\gamma})}{\sigma}\right)^2\}}{\sqrt{2\pi}\sigma y_{i}\Phi\left(\frac{\log(y_{mo}(\bold{x_i};\boldsymbol{\gamma}))-\mu(\bold{x_i};\boldsymbol{\gamma})}{\sigma}\right)}\right)\right.\\ \nonumber
		& \quad \quad  \left. \qquad \mathbb{I}[y_{i} < y_{mo}(\bold{x_i};\boldsymbol{\gamma})] 
		+  \ln	(1-r_{LN,T}(\bold{x_i};\boldsymbol{\gamma})) \mathbb{I}[y_{i} \ge y_{mo}(\bold{x_i};\boldsymbol{\gamma})] \right. \\ 
		\quad \quad \qquad & \quad \quad \quad 		\left.   +\ln\left(\frac{f_{T}(y_i)}{1-F_{T}(y_{mo}(\bold{x_i};\boldsymbol{\gamma}))}\right) \mathbb{I}[y_{i} \ge y_{mo}(\bold{x_i};\boldsymbol{\gamma})]\right].
	\end{align}
	Where $r_{LN,T}(\bold{x_i};\boldsymbol{\gamma})$ is the covariate dependent mixing weight linked to the head part of the composite regression model with Lognormal distribution for the head and $T$ distribution for the tail. $y_{mo}(\bold{x_i};\boldsymbol{\gamma})$ is the covariate dependent varying threshold point.
	All of the parameters ($\boldsymbol{\Theta}$) of Mode-Matching composite Lognormal-$T$ regression model are estimated using the numerical optimization tool \texttt{optim()}, included in the \textbf{stats} package of \texttt{R} programming language. The \texttt{BFGS} $(Broyden-Fletcher-Goldfarb-Shanno)$ algorithm passed to \texttt{optim()} via the argument \texttt{method}, is used for maximization. The maximum likelihood estimation  procedure can be applied relatively easy via transformation-back transformation approach given in \cite{11} as compared to augmented Lagrange multiplier method given in \cite{19}. Some of the parameters involved in the estimation procedure are subject to constraints. A transformation/back transformation strategy has been applied in order to make the maximization of $l(\boldsymbol{\Theta}|y_{i})$ unconstrained, as required by the \texttt{BFGS} algorithm. Particularly, the original constrained parameters are transformed  to unconstrained real values and thereafter the log-likelihood $l(\boldsymbol{\Theta}|y_{i})$ is maximized with respect to the unconstrained parameters. To acquire the original constrained parameter estimates, a back-transformation is used. Table \ref{TBM} shows the necessary transformation and back transformations used in the ML estimation procedure for the parameters of the composite Lognormal-$T$  regression models.  
	
	% \usepackage{multirow}
	\begin{table}[H]
		\centering
		\caption{Transformation/ back transformation used for the constrained parameters of the composite Lognormal-$T$ regression model}
		\begin{tabular}{lll} \hline \hline
			\multicolumn{1}{l}{Model}         & \multicolumn{2}{l}{Transformation   / Back transformation} \\ \hline \hline
			\multirow{3}{*}{Lognormal-Burr}   & \multicolumn{2}{l}{$\Tilde{\sigma}$= $\ln (\sigma)$  $\longleftrightarrow$ $\sigma= \exp(\Tilde{\sigma})$}                                     \\
			& \multicolumn{2}{l}{$\Tilde{\alpha}$= $\ln (\alpha-1)$  $\longleftrightarrow$ $\alpha= \exp(\Tilde{\alpha})+1$}                                     \\
			& \multicolumn{2}{l}{$\Tilde{\delta}$= $\ln (\delta)$  $\longleftrightarrow$ $\delta= \exp(\Tilde{\delta})$}                                     \\ \hline 
			
			\multirow{3}{*}{Lognormal-Stoppa} & \multicolumn{2}{l}{$\Tilde{\sigma}$= $\ln (\sigma)$  $\longleftrightarrow$ $\sigma= \exp(\Tilde{\sigma})$}                                   \\
			& \multicolumn{2}{l}{$\Tilde{\alpha}$= $\ln (\alpha-1)$  $\longleftrightarrow$ $\alpha= \exp(\Tilde{\alpha})+1$}                                   \\
			& \multicolumn{2}{l}{$\Tilde{\delta}$= $\ln (\delta)$  $\longleftrightarrow$ $\delta= \exp(\Tilde{\delta})$}     \\ \hline
			
			\multirow{2}{*}{Lognormal-GlogM}  & \multicolumn{2}{l}{$\Tilde{\sigma}$= $\ln (\sigma)$  $\longleftrightarrow$ $\sigma= \exp(\Tilde{\sigma})$}                                    \\
			& \multicolumn{2}{l}{$\Tilde{\alpha}$= $\ln (\alpha)$  $\longleftrightarrow$ $\alpha= \exp(\Tilde{\alpha})$}                                    \\ \hline \hline                            
		\end{tabular}
		\label{TBM}
	\end{table}

	\subsection{Calculation of standard errors}
	The Hessian of the log-likelihood function at the maximum likelihood estimate was computed using numerical tools. This Hessian, also known as the observed Fisher information matrix, can be used to determine standard errors of parameter estimates. $I(\boldsymbol{\Theta})=-\partial^{2}l(\boldsymbol{\Theta})/\partial{\boldsymbol{\Theta}}\partial{\boldsymbol{\Theta}^{'}}$. The inverse of this $I(\boldsymbol{\Theta})$ provides estimate for covariance matrix, $\text{Cov}(\boldsymbol{\Theta})$. The estimates of the standard error for the parameter estimates can be calculated by taking square root of the diagonal elements of the covariance matrix. 
	
	\section{Empirical investigation and results based on Real Data} \label{RDA}
	In this section, we illustrate the proposed methodology using the motor third-party liability (\texttt{MTPL}) insurance policies with non-zero property claims for the years 2012 to 2019. A major insurance firm in Greece generously provided the data set for this study. The data set contains 7263 policies of motor vehicle insurance collected between 2012 and 2019 which have complete records. The following section provides a detailed description of the (\texttt{MTPL}) data set. The response variable associated with the \texttt{MTPL} data set is \texttt{tcost bi} i.e. the cost of a bodily injury claim is a numeric vector showing the total amount of bodily injury claims. The covariates information that we incorporated in the model fitting is the car cubism (CC), the Policy Type, the Vehicle age, and the MTPL cost.
	
\begin{figure}[H]
		\centering
		\includegraphics[width=10.5cm]{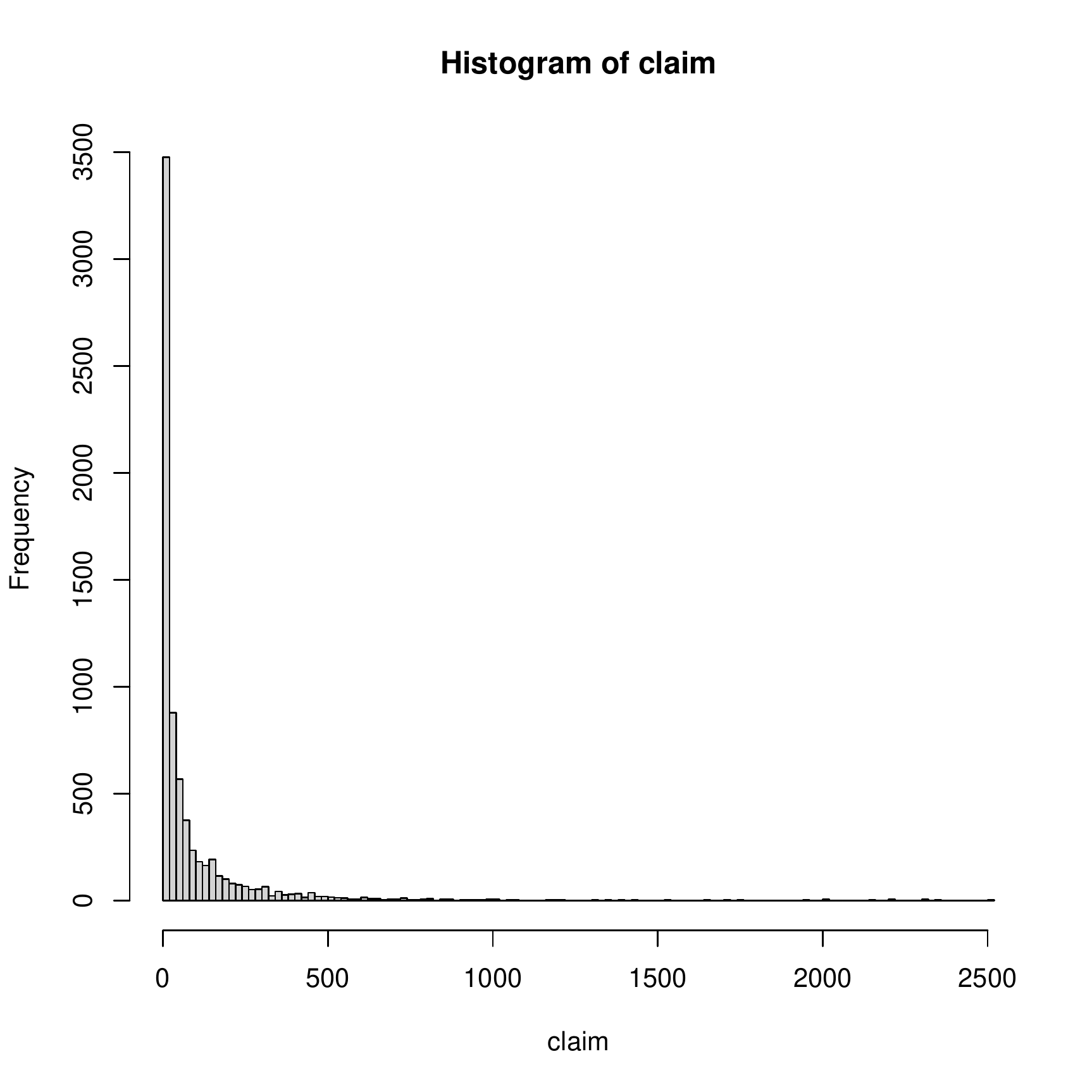}
		\caption{Histogram of the bodily injury claim for the \texttt{MTPL} data set}
		\label{histclaim}
	\end{figure}

\begin{figure}[H]
		\centering
		\includegraphics[width=10cm]{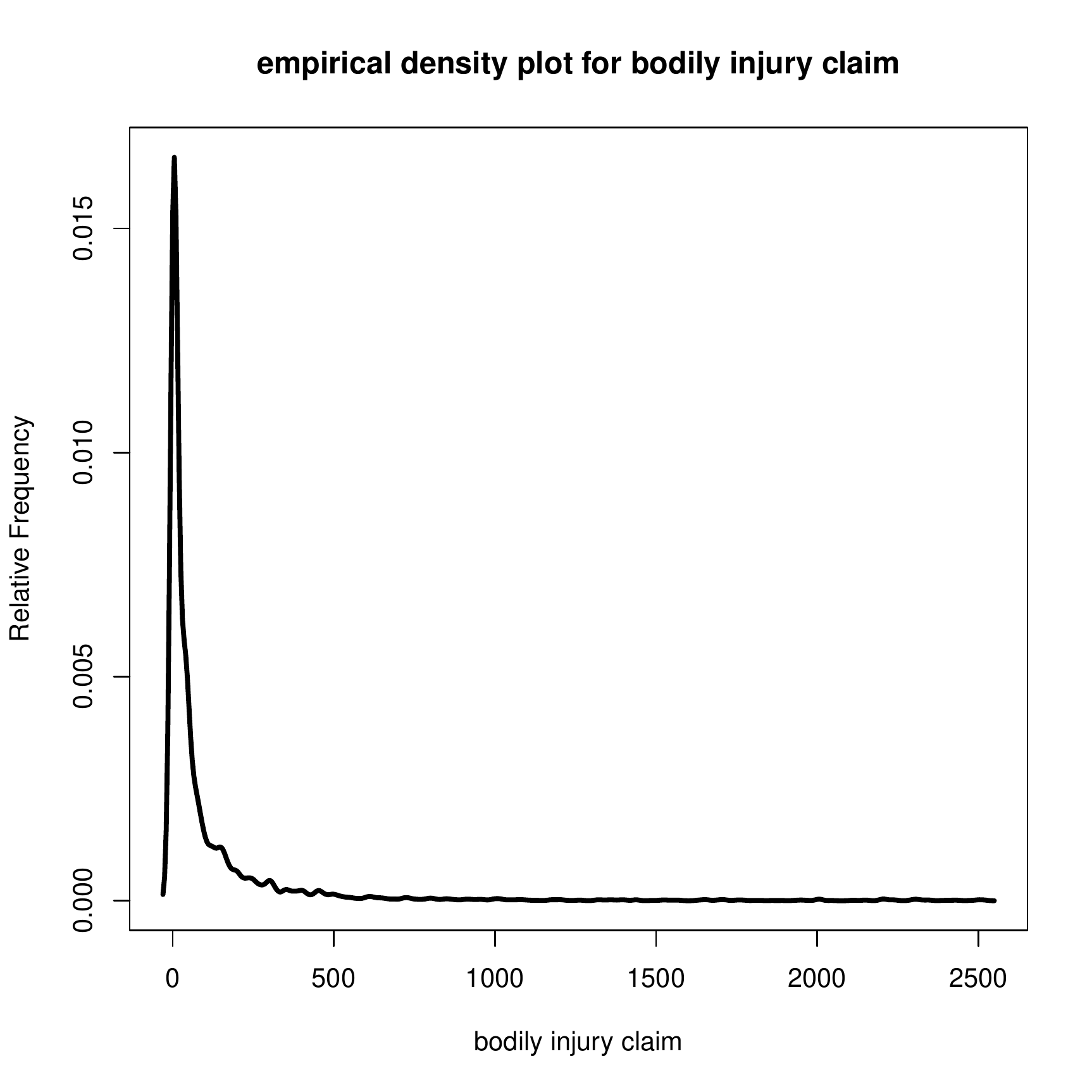}
		\caption{Empirical density of claim}
		\label{epdfclaim}
	\end{figure}

\begin{figure}[H]
	\centering
		\includegraphics[width=16cm]{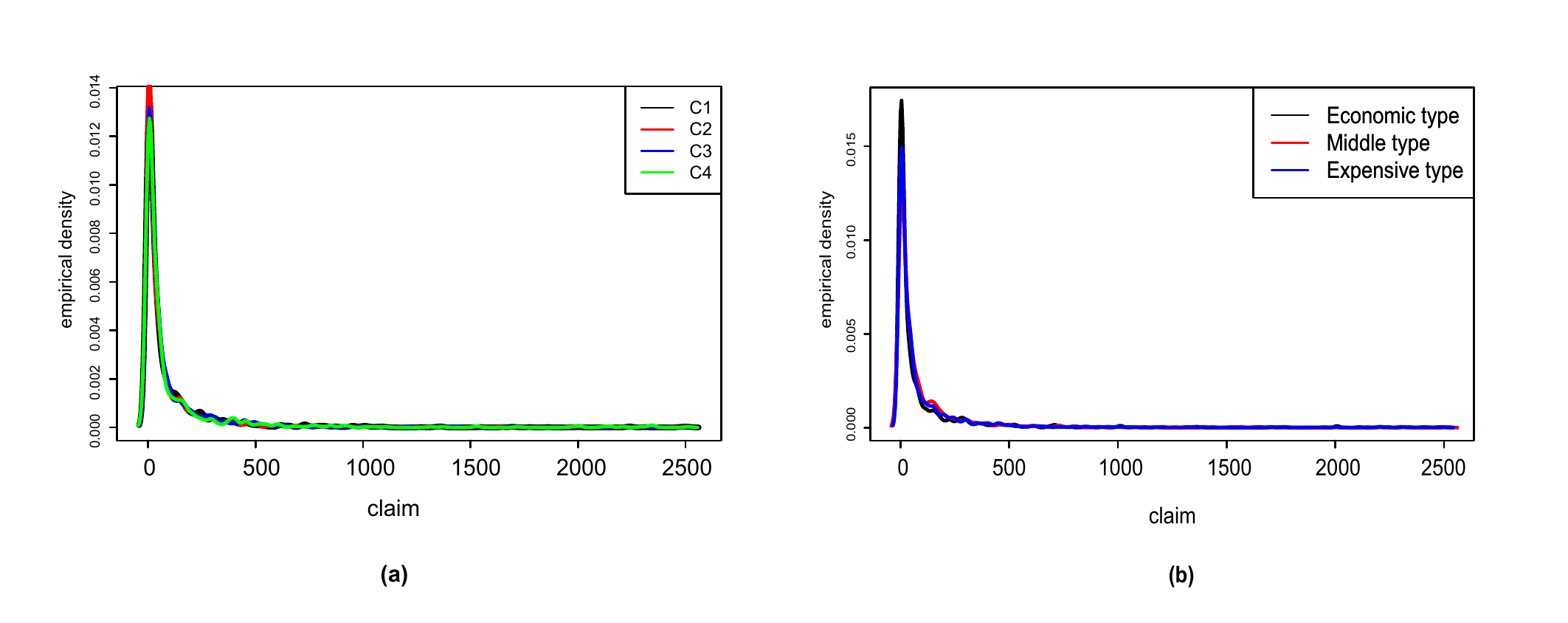}
		\caption{Empirical density plots of bodily injury claims for various subcategories of (a) car cubism (CC) covariate and (b) Policy Type covariate}
	\label{epdfc}
\end{figure}

	\begin{itemize}
		\item \texttt{CC} consists of  car cubism with four categories C1 (0-1299 cc), C2 (1300-1399 cc), C3 (1400-1599 cc) and C4 (Equal or Greater than 1600 cc).
		\item The explanatory variable \texttt{Policy Type} contains three categories C1 (Economic type), C2 (Middle type) and C3 (Expensive type) .
		\item  \texttt{Vehicle age} consists 0 - 38 years old vehicles.
	\end{itemize}
	Figure \ref{histclaim} contains the histogram for the response variable \texttt{tcost bi}. The skewness for response variable \texttt{tcost bi} of  \texttt{MTPL} dataset is 5.31 which results in  positive skewness in the data. The difference between \texttt{$Q_{3}$} and maximum value of response variable shows the heavy-tailed behavior of the \texttt{MTPL} dataset. Table \ref{srce} presents the descriptive statistics for response variable  \texttt{tcost bi}  and continuous explanatory variables available with the response variable. Table \ref{scev} gives the frequency distribution of the categorical explanatory variables of the \texttt{MTPL} data set. 
	\begin{table}[H]
		\centering
		\caption{Summary of response variable and continuous explanatory variables of the \texttt{MTPL} data set.}
		\scalebox{1.01}{
			\begin{tabular}{ccccccccc} \hline \hline
				Variable & Minimum & Maximum & \texttt{Q1} & Median & \texttt{Q3}& Mean & Skewness & Kurtosis \\ \hline \hline
				\texttt{tcost bi}& 0.009&2519.58&3.18&24.13& 91.21&110.17&5.31&36.88     \\
				\texttt{Vehicle Age} & 0.00 & 38 & 8 & 12 & 16 & 12.17 & 0.39 & 3.69   \\
				\texttt{MTPL cost}&0.02& 256.01 & 1.62 & 4.67 & 12.01 & 12.88 & 5.23 & 36.08 \\ \hline
		\end{tabular}}
		\label{srce}
	\end{table}
	\begin{table}[H]
		\centering
		\caption{Summary of categorical explanatory variables of the \texttt{MTPL} data set.}
		\scalebox{0.85}{
			\begin{tabular}{|c|c|} \hline \hline
				CC & Policy Type  \\ \hline \hline
				C1 : 2036   & C1 : 1144      \\
				C2 : 2417   &C2 : 1940     \\
				C3 : 1833    &C3 : 4179     \\
				C4 : 0977   &--    \\  \hline                                             
		\end{tabular}}
		\label{scev}
	\end{table}
	Figure \ref{epdfclaim} represents an empirical density plot for the bodily injury claim which indicates the heavy-tailed and uni-modal nature of the data. The empirical density plots for the bodily injury claim against various sub-categories of covariates are displayed in Figure \ref{epdfc}. We can see that some categorical covariates appear to have an impact on the distributional modality of the \texttt{MTPL} data set. The \texttt{Policy Type} covariate with “Middle Type” category has a slightly higher probability assigned to the tail part as compared to the remaining levels of \texttt{Policy Type} covariate. Different variables impact different parts of the distribution  (e.g., covariates can affect mixing weight $r$, tail-heaviness as well as modal part of the \texttt{MTPL} data set), a composite regression model must have sufficient flexibility to model distributional modality, mixing weight $r$ and tail-heaviness of the distribution. The flexibility in the modeling of different parts of the distribution may be achieved by introducing covariates in the scale parameter which is a part of the analytical expression of the mode of the Mode-Matching composite regression model. The covariate dependent mode is the threshold point of the composite regression model which results in the varying threshold across the individual claimant's risk behavior.
	
\subsection{Model Comparison}
	In this section, we present the results based on the three model selection criterion for the proposed models. Table \ref{NLL} provides the values negative of the maximum of the log-likelihood (NLL), Akaike’s information criterion (AIC) and Bayesian information criterion (BIC). Note that for all three measures of model validation, smaller values indicate a better fit of the model to the empirical data. The formulae involved in the computation of above-mentioned model selection criterion are\\
	The AIC can be computed as
	\begin{equation*}
		AIC=-2{l}(\hat{\boldsymbol{\Theta}})+ 2 \times df,
	\end{equation*} 
	where ${l(\hat{\boldsymbol{\Theta}})}$ is the maximum of the log-likelihood and $\hat{\boldsymbol{\Theta}}$ is the vector of the estimated model parameters. 
	and the BIC is given by
	\begin{equation*}
		BIC=-2{l}(\hat{\boldsymbol{\Theta}})+ \text{log}(n) \times df,
	\end{equation*}
	where $n$ is sample size of the data set and $df$ is the number of fitted parameters of the model.
	
	To show the effect of considering varying threshold in comparison with fixed threshold, we fitted the Lognormal-Burr, Lognormal-Stoppa and Lognormal-GlogM composite regression with a fixed threshold. The NLL corresponding to the composite Lognormal-Burr regression model is 32436.11 and that of composite Lognormal-Stoppa and composite Lognormal-GlogM regression models are  32094.25 and 32225.22 respectively. The following results from Table \ref{NLL} show that the composite Lognormal-Burr regression model with varying threshold performs better than the remaining composite regression models with varying threshold followed by the composite Lognormal-Stoppa regression model.
	
	\begin{table}[H]
		\centering
		\caption{Values of Negative Log-Likelihood Function, AIC and BIC for \texttt{MTPL} dataset }
		\begin{tabular}{lcccc} \hline \hline
			Model         & df & NLL           & AIC      & BIC      \\ \hline \hline
			Lognormal-Burr & 11 & 31273.95		 &	62569.91		 & 62590.37	 \\
			Lognormal-Stoppa & 11 & 31594.83		 &	63211.66		 &	63232.13	 \\ 
			Lognormal-GlogM  & 10  & 31681.67		  & 63383.34		 & 63401.95 \\ \hline \hline
		\end{tabular} 
		\label{NLL}
	\end{table}
	
	\subsection{Modelling  Results}
	Making use of the additional information, we aim to better explain the total losses in terms of the set of covariates by using the composite Lognormal-$T$  regression model.
	The results for the parameter estimates and standard errors (SE) are calculated for the composite Lognormal-$T$ regression and presented in the Table \ref{paraest}.
	
	\begin{table}[H]
		\centering
		\caption{Parameter estimates and standard error (SE) for the composite regression models}
		\begin{tabular}{l|cccccc} \hline \hline
			\multirow{3}{*}{Covariates} & \multicolumn{6}{c}{Model}                                                                  \\ \hline
			& \multicolumn{2}{c}{LN-Burr} & \multicolumn{2}{c}{LN-Stoppa} & \multicolumn{2}{c}{LN-GlogM} \\ \hline \hline
			& Estimate      & SE          & Estimate       & SE           & Estimate       & SE          \\ \hline \hline
			\texttt{Intercept}                          & -0.7839       & 0.1265     & -6.4068         & 0.9101       & -0.3373         &0.1304     \\
			\texttt{CC : 2}                          &  -1.2333       & 0.0771     & 1.8395       & 0.0855       & 1.9041       & 0.0935     \\
			\texttt{CC : 3}                          &    -0.9538    & 0.0732      & 1.3518        & 0.0853      & 1.5058       & 0.0833     \\
			\texttt{CC : 4}                           & -0.4102       & 0.0815      & 0.9369       & 0.0922       & 0.9514         & 0.0948     \\
			\texttt{Policy-Type : Middle}                           &  -1.5709    & 0.0641      & 2.0928        & 0.0619       & 2.3234          & 0.0796     \\
			\texttt{Policy-Type : Expensive}                          &   -1.8162     & 0.1974      & 2.8798       & 0.2069       & 2.8867        & 0.1981     \\
			\texttt{Vehicle Age}                            &  0.0287     & 0.0041      & -0.0597        & 0.0045       & -0.0449        & 0.0047     \\
			\texttt{MTPL Cost}                            &  -0.0303       & 0.0008      & 0.0257       & 0.0005      & 0.0233        &0.0004     \\
			$\sigma$                       & 8.2441       & 0.0001      & 7.6706         & 0.0008       & 7.8837         & 0.0003     \\
			$\alpha$                       & 1.0677       & 0.2951      & 106.3263        & 0.6549       & 0.7399        & 0.0233      \\
			$\delta$                       & 0.8771       & 0.03251      & 0.7235       & 0.0185       & --             & --    \\ \hline \hline     
		\end{tabular}
		\label{paraest}
	\end{table}
	
	\subsubsection{Diagnostic results}
	Further we test the significance of the individual regression coefficient as $\mathscr{H}_0 : \gamma_s = 0$ against $\mathscr{H}_1 : \gamma_s \ne 0$ for $s = 1,\dots p$. For testing the hypotheis, the $t$-ratio corresponding to each regression coefficient can be given by $t-\text{ratio} = \frac{\hat{\gamma_s}}{SE(\hat{\gamma_s})}$. We reject the null hypothesis if |$t$-ratio| > $t_{n-p, 1-\frac{\alpha}{2}}$, where $p$ is the total number of regression coefficients in the model and $\alpha$ is the level of significance. Most of the levels of available covariates are showing high impact on both distributional modality as well as thicker right tail of the \texttt{MTPL} data set. Table (\ref{tpvalue}) facilitate the diagnostic results for the composite regression models. 
	
	\begin{table}[H]
		\centering
		\caption{Diagnostic results for the composite regression models}
		\begin{tabular}{l|cccccc} \hline \hline
			\multirow{3}{*}{Covariates} & \multicolumn{6}{c}{Model}                                                                  \\ \hline
			& \multicolumn{2}{c}{LN-Burr} & \multicolumn{2}{c}{LN-Stoppa} & \multicolumn{2}{c}{LN-GlogM} \\ \hline \hline
			& $t$-ratio  & $p$-Value    & $t$-ratio  & $p$-Value     & $t$-ratio  & $p$-Value          \\ \hline \hline
			\texttt{Intercept}          & -6.1969       &  $\le 0.001$      & -7.0396        &  $\le 0.001$       & -2.5866        & $\le 0.001$     \\
			\texttt{CC : 2}             &  -15.9961      &  $\le 0.001$      & 21.5146      &  $\le 0.001$        & 20.3647       & $\le 0.001$     \\
			\texttt{CC : 3}          &    -13.0301    &  $\le 0.001$      & 15.8475       &  $\le 0.001$      & 18.0768      &  $\le 0.001$      \\
			\texttt{CC : 4}          & -5.0331       & $\le 0.001$       & 10.1616      &  $\le 0.001$        & 10.0358         &  $\le 0.001$     \\
			\texttt{Policy-Type : Middle}     &  -24.5071    &  $\le 0.001$       & 33.8093        &  $\le 0.001$        & 30.2132         &  $\le 0.001$     \\
			\texttt{Policy-Type : Expensive} &   -9.2006    &  $\le 0.001$       & 13.9188       &  $\le 0.001$        & 14.5719        &  $\le 0.001$     \\
			\texttt{Vehicle Age}     &  7.0001     &  $\le 0.001$       & -13.2666        &  $\le 0.001$        & -9.5531        &  $\le 0.001$    \\
			\texttt{MTPL Cost}       &  -37.8571      &  $\le 0.001$       &51.4001       &  $\le 0.001$       & 58.25      & $\le 0.001$     \\ \hline \hline     
			
		\end{tabular}
		\caption*{*All the explanatory variables are statistically significant at 5\% level of significance}
		\label{tpvalue}
	\end{table}
	
	\subsubsection{Normalized Quantile Residual}
\begin{figure} [H]
		\centering
		\includegraphics[width=5cm]{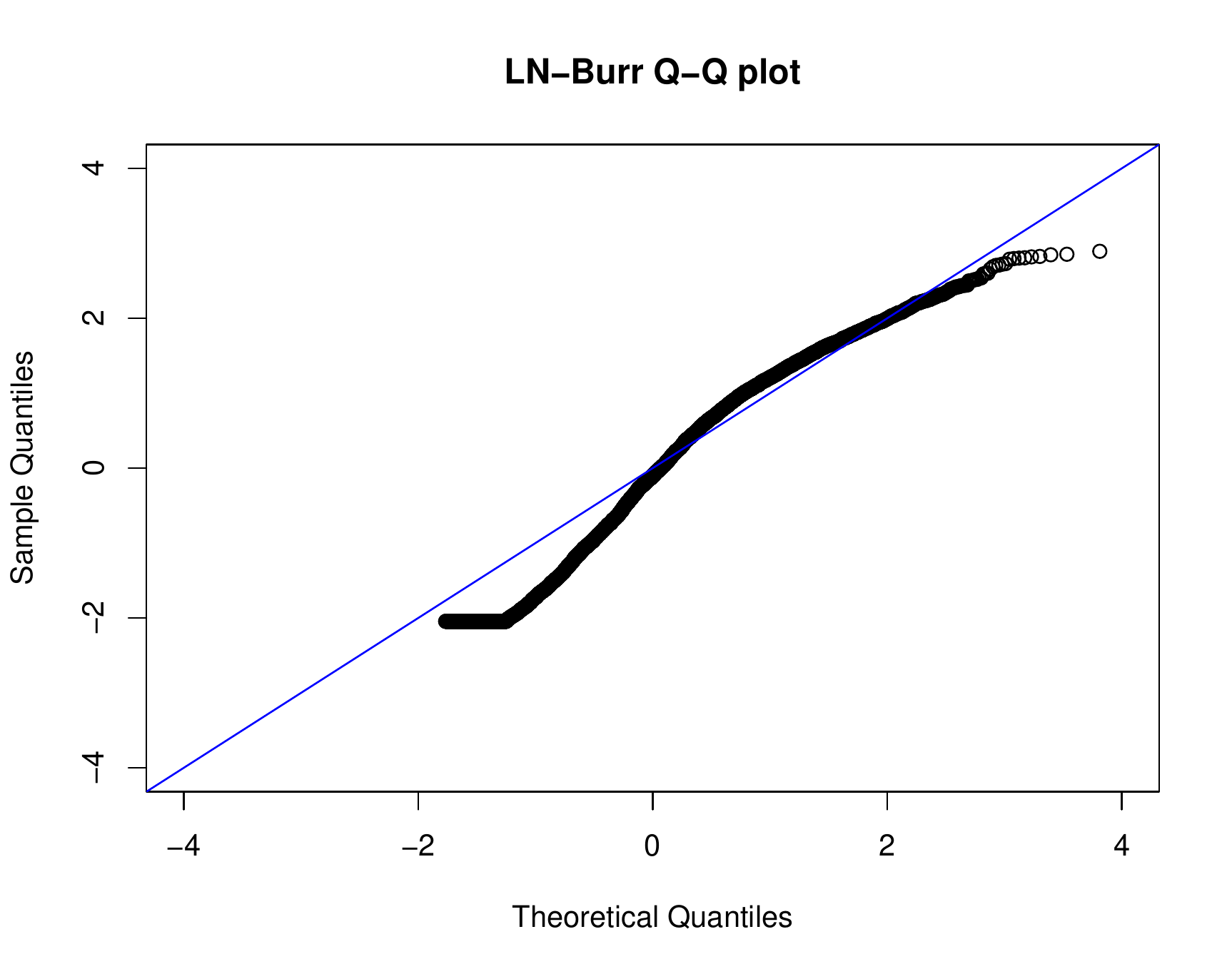}
		\includegraphics[width=5cm]{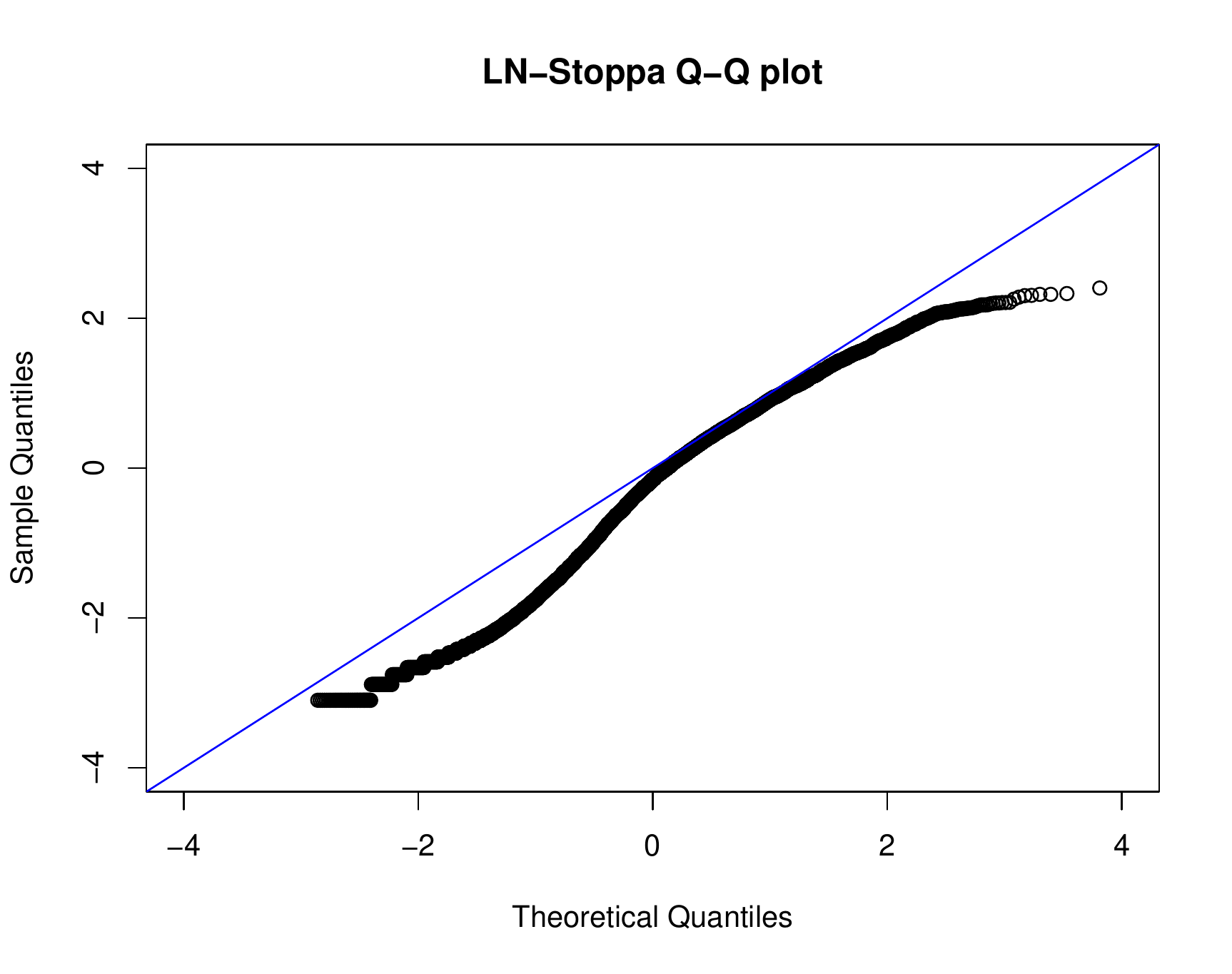}
		\includegraphics[width=4.9cm]{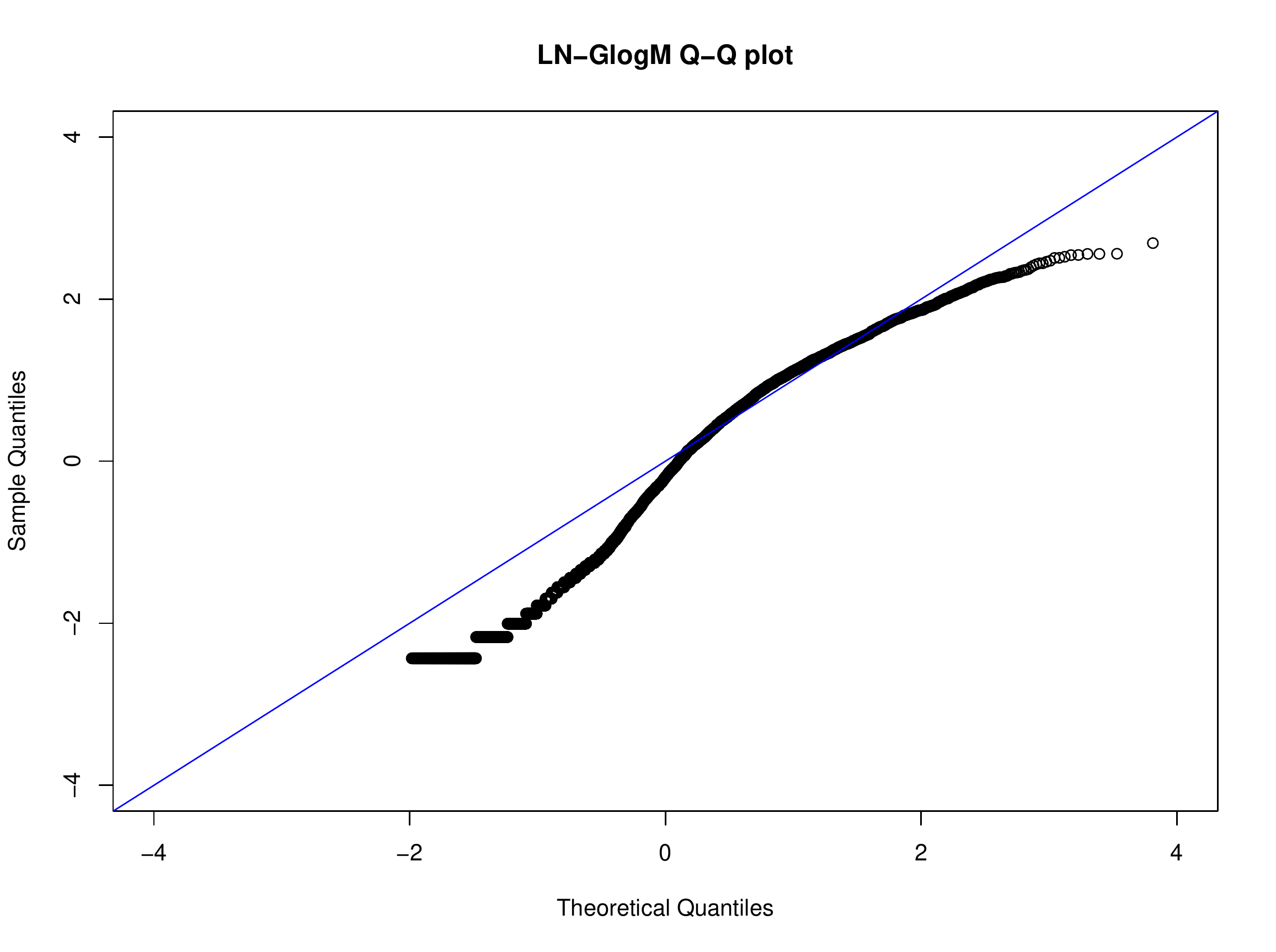}
		
		\caption{$QQ$-Plot for the proposed composite regression models}
		\label{qq}
	\end{figure}

	To test the adequacy of fitting of the proposed composite regression models, we provide the $QQ$-plots of the normalized quantile residuals. The formula involved in the computation of normalized quantile residuals is as follows
	\begin{equation}
		k_i=\Phi^{-1}(F(y_i;\hat{\boldsymbol{\Theta}})).
	\end{equation}
	Where $i=1,2,\cdots,n$, $\hat{\boldsymbol{\Theta}}$ represents the vector of the estimated model parameters, $F(y_i;\hat{\boldsymbol{\Theta}})$ is the cdf of the composite regression model evaluated at $y_i$ using $\hat{\boldsymbol{\Theta}}$ and $\Phi^{-1}$ denotes the quantile function of the standard Normal distribution. Figure \ref{qq} gives the $QQ$-Plots corresponding to the Lognormal-Burr, the Lognormal-Stoppa and the Lognormal-GlogM composite regression models. The composite Lognormal-Burr gives the better fit for the tail part of the \texttt{MTPL} data as compared to remaining composite regression models.
	
	\section{Conclusions} \label{CONC}
	In this paper, we propose a new family of composite regression models using the Mode-Matching procedure. Regression analysis for the heavy-tailed response variable is becoming a new avenue of research where the usual regression models fail to capture the tail of the data set using available covariates. The main purpose of this paper is to introduce a Mode-Matching composite Lognormal-$T$  regression model which can efficiently capture the influence of covariates on the tail behavior as well as the modal part of the claim severity data set. In this study, we provide the detailed methodology involved in the generation of three composite regression models developed using the Mode-Matching procedure, namely the Lognormal-Burr, Lognormal-GlogM and Lognormal-Stoppa composite regression models. We propose to incorporate   available covariates in the scale parameter of the tail part of the composite Lognormal-$T$  regression model. The computational challenges that may arise in the model estimation procedure are also discussed to efficiently estimate the model parameters. One real world insurance data set namely, third party liability data set (\texttt{MTPL}) is exemplified to show the applicability of our proposed models and estimation procedure. The model comparison and modeling results are presented to show the need for improvement in the composite model by incorporating the covariates in the model framework. The findings demonstrate that the composite Lognormal-Burr regression model outperforms alternative heavy-tailed regression models.
	
	\section*{Funding}
	The present work is a part of a project granted by the Department of Science \& Technology, Government of India under the Core Research Grant scheme (CRG/2019/002993).  D.B. and G.A. thank the funding agency for financial support.


\begin{thebibliography}{}
		\bibitem{1} C.J. Adcock , M. Eling, and N. Loperfido, \textit{Skewed distributions in finance and actuarial science: a review}, The European Journal of Finance, 21 (2015), pp. 1253-1281.
	
	\bibitem{2} S. Ahn, J.H. Kim, and V. Ramaswami, \textit{A new class of models for heavy tailed distributions in nance and insurance risk}, Insurance: Mathematics and Economics, 51 (2012), pp. 43-52.
	
	\bibitem{3} H. Akaike, \textit{A new look at the statistical model identification}, IEEE Transactions on Automatic Control 19 (1974), pp. 716–723. 
	
	\bibitem{4} S.A. A. Bakar, N. A. Hamzaha,  M. Maghsoudia,  and  S. Nadarajah, \textit{Modeling loss data using composite models}, Insurance: Mathematics and Economics, 61 (2015), pp. 146–154.
	
	\bibitem{5} G. Bettina, and M. Tatjana, \textit{Extending composite loss models using a general framework of advanced computational tools}, Scandinavian Actuarial Journal, 8 (2019), pp. 642-660.
	
	\bibitem{6} D. Bhati, E. Calderín-Oj\'eda, and M. Meenakshi, \textit{A New Heavy Tailed Class of Distributions Which Includes the Pareto}, Risks, 7(2019), pp. 99.
	
	\bibitem{7} D. Bhati, S. Ravi, \textit{On generalized log-Moyal distribution: A new heavy tailed size distribution}. Insurance: Mathematics and Economics, 79 (2018), pp. 247-259.
	
	
	\bibitem{8} E. Calderín-Oj\'eda,  and C. F. Kwok, \textit{Modeling claims data with composite Stoppa models}, Scandinavian Actuarial Journal, 2016 (2016), pp. 817–836. 
	
	\bibitem{9} R. Ciumara, \textit{An actuarial model based on the composite Weibull-Pareto distribution}, Mathematical Reports- Bucharest, 8 (2006) , pp. 401–414.
	
	\bibitem{10} K. Cooray, and M. Ananda,  \textit{Modeling actuarial data with a composite lognormal-Pareto model}, Scandinavian Actuarial Journal, 2005 (2005), pp. 321–334.
	
	\bibitem{11} S. D. Tomarchio, and A. Punzo, \textit{Dichotomous unimodal compound models: application to the distribution of insurance losses}. Journal of Applied Statistics, 47 (2020), pp. 2328-2353.
	
	\bibitem{12} M. Eling, \textit{Fitting insurance claims to skewed distributions: Are the skew-normal	and skew-student good models?}, Insurance: Mathematics and Economics, 51 (2012), pp. 239-248.
	
	\bibitem{13} T. C. Fung, G. Tzougas, and M. Wuthrich,  \textit{Mixture composite regression models with multi-type feature selection}, preprint (2021). Available at arXiv preprint arXiv:2103.07200.
	
	\bibitem{14} G. Gan, and E.A. Valdez, \textit{Fat-Tailed Regression Modeling with Spliced Distributions}, North American Actuarial Journal, 22 (2018), pp. 554-573.
	
	
	\bibitem{15} C. Kahadawala, \textit{The Weibull–Pareto Composite Family with Applications to the Analysis of Unimodal Failure Rate Data}, Communications in Statistics—Theory and Methods, 38 (2009), pp. 1901-1915.
	
	\bibitem{16} R. Kazemi, and M. Noorizadeh, \textit{A comparison between skew-logistic and skewnormal distributions}, Matematika, 31 (2015), pp. 15-24.
	
	\bibitem{17} S. A. Klugman, H. H. Panjer, and G. E. Willmot, \textit{Loss models: from data to decisions}, John Wiley \& Sons (2012).
	
	\bibitem{18} H. Konşuk \"{u}nl\"{u}, \textit{A New Composite Lognormal-Pareto Type II Regression Model to Analyze Household Budget Data via Particle Swarm Optimization}, Soft Computing, 26 (2021), pp. 2391-2408.
	
	
	\bibitem{19} Z. Li, F. Wang, and Z. Zhao, \textit{ A new class of composite GBII regression models with varying threshold for modelling heavy-tailed data}, preprint (2022). Available at arXiv preprint arXiv:2203.11469.
	
	\bibitem{20} T. Miljkovic,  and  B. Grün, \textit{ Modeling loss data using mixtures of distributions}, Insurance: Mathematics and Economics, 70 (2016), pp. 387–396.
	
	
	\bibitem{21} S. Nadarajah,  and  S. Bakar, \textit{New composite models for the Danish fire insurance data}, Scandinavian Actuarial Journal, 2014 (2014), pp. 180–187.
	
	\bibitem{22} M. Pigeon,  and  M. Denuit,  \textit{ Composite lognormal-Pareto model with random threshold}, Scandinavian Actuarial Journal, 2011 (2011), pp. 177–192.
	
	\bibitem{23} A. Punzo, L. Bagnato, and A.  Maruotti, \textit{ Compound unimodal distributions for insurance losses}, Insurance: Mathematics and Economics, 81 (2018), pp. 95-107.
	
	
	\bibitem{24} D. Scollnik,   \textit{On composite lognormal-Pareto models}, Scandinavian Actuarial Journal, 2007 (2007), pp. 20–33.
	
	\bibitem{25} D. P. Scollnik,  and C. Sun, \textit{Modeling with Weibull-Pareto models}, North American Actuarial Journal, 16 (2012), pp. 260–272. 
	
	\bibitem{26} G. Stoppa, \textit{Proprieta campionarie di un nuovo modello Pareto generalizzato} [Sample properties of a new generalized Pareto model],  Atti XXXV Riunione Scientifica della Societa Italiana di Statistica, Padova: Cedam, 35(1990), pp. 137-144.
	
	\bibitem{27} G. Tzougas, and D. Karlis, \textit{ An EM algorithm for fitting a new class of mixed exponential regression models with varying dispersion}, ASTIN Bulletin: The Journal of the IAA, 50(2020), pp. 555-583.
	
	\bibitem{28} R. Verbelen, L. Gong,  K. Antonio, A. Badescu,  and S. Lin,  \textit{Fitting mixtures of Erlangs to censored and truncated data using the EM algorithm}, ASTIN Bulletin, 45(2015), pp. 729-758.
	
	\bibitem{29} R. Vernic, \textit{Multivariate skew-normal distributions with applications in insurance}, Insurance: Mathematics and Economics, 38 (2006), pp. 413-426.
		
		
	\end{thebibliography}
\end{document}